\documentclass[11pt,a4paper,floatfix]{article}
\usepackage{authblk}
\usepackage{geometry}
\usepackage[utf8]{inputenc}

\usepackage{amsmath}
\usepackage{amssymb}
\usepackage{graphicx}
\usepackage{import}
\usepackage{xcolor}
\usepackage[linkcolor=black,
colorlinks=true,
citecolor=black,
urlcolor=cyan]{hyperref}
\usepackage{cleveref}
\crefformat{equation}{#2EQ.~#1#3}
\crefformat{figure}{#2FIG.~#1#3}
\crefformat{table}{#2TAB.~#1#3}
\crefformat{section}{#2SEC.~#1#3}
\usepackage[normalem]{ulem}

\usepackage{physics}
\newcommand{\iu}{{i\mkern1mu}}

\begin{document}
\title{
Revealing the coaction of viscous and multistability hysteresis in an adhesive, nominally flat  punch: A combined numerical and experimental study
}

\author[1,2]{Christian M\"uller}
\author[1]{Manar Samri}
\author[1]{René Hensel}
\author[1,2]{Eduard Arzt}
\author[1,2]{Martin H. M\"user}
\affil[1]{INM - Leibniz Institute for New Materials}
\affil[2]{Dept. of Materials Science, Saarland University}



\date{\today}

\begin{abstract} 


Viscoelasticity is well known to cause significant hysteresis of crack closure and opening when an elastomer is brought in and out of contact with a flat, rigid, adhesive counterface.
{A separate origin of adhesive hysteresis is  small-scale, elastic multistability.}
Here, we study a system in which both mechanisms act concurrently.
Specifically, we compare the simulated and experimentally measured time evolution of the interfacial force and the real contact area between a soft elastomer and a rigid, flat punch, to which small-scale, single-sinusoidal roughness is added. 
To this end, we further the Green's function molecular dynamics method and extend 
recently developed imaging techniques to elucidate the rate- and preload-dependence of the  pull-off process.
%
%
Our results reveal that hysteresis is much enhanced when the saddle points of the topography come into contact, which, however, is impeded by viscoelastic forces and may require sufficiently large preloads. 
A similar coaction of viscous- and multistability effects is expected to occur in macroscopic polymer contacts and to be relevant, e.g., for pressure-sensitive adhesives and modern adhesive gripping devices.
%
\end{abstract}

\maketitle

\section{Introduction}

Bringing two surfaces into contact and separating them again is generally associated with a net, rate-dependent energy loss.
Several processes can cause this hysteresis to occur, in particular, physicochemical interfacial aging~\cite{Chen1991JPC,Liu2012PRL}, such as chain interdigitation in polymer-polymer contacts~\cite{Maeda2002S}, viscoelastic relaxation in the vicinity of and far from true contact~\cite{Giri2001L,Shull2002MSER,Lorenz2013JPCM,Tiwari2017SM}, and the formation of capillaries~\cite{Pickering2001JAST,Feiler2006L,Israelachvili2011}, to name a few.
%
{Over the years}, elastic multistability~\cite{Prandtl1928ZAMM,Tomlinson1929LEDPMJS} has also received much attention as a potential adhesive dissipation mechanism occuring during the relative motion of nominally flat surfaces, i.e., the discontinuous jump of small-scale asperities in and out of contact~\cite{Tomlinson1929LEDPMJS,Thomson1971JAP,Gao1989JAMTA,Zheng2004CP,Guduru2007JMPS,Glassmaker2007PNASUSA,Kesari2010PML,Xia2012PRL,Carbone2015PRE,Dalvi2019PNASUSA,Wang2021L} during quasi-static motion, or the discontinuous motion of a contact line during approach and retraction resulting from chemical or structural surface heterogeneity~\cite{Sanner2022JMPS}. 

Ascertaining what adhesion-hysteresis mechanism dominates under what circumstances is a difficult task, because analytical solutions for the rate- and/or the preload dependence of the pull-off force scarcely exist, even when only one relaxation process dominates. 
Moreover, it is certainly conceivable that competing mechanisms, e.g., contact aging and contact growth, lead to a similar, for example, logarithmic time dependence {of} the pull-off force {on} the waiting time.
The validity of models and theories, irrespective of whether they are solved analytically or numerically, should therefore be tested against information additional to load-displacement relations and their dependence on rate, waiting time, and preload.
A central quantity to be known is the time evolution of true contact, including its size and shape. \par

While small-scale features of adhesive experimental and \textit{in-silico} contacts have been successfully compared in the recent past, such as in the contact-mechanics challenge~\cite{Mueser2017TL,Bennett2017TL} or to demonstrate the breakdown of Amonton's law at the small scale in soft-matter contacts~\cite{Weber2018NC}, we are not aware of 
{similarly sophisticated}
studies involving time-dependent phenomena as they occur during adhesion hysteresis.
Detailed comparisons between simulations and experiments {are often} conducted only during compression but not during retraction. 
This may be the case because  simultaneously simulating  multi-scale roughness, viscoelasticity, and adhesion has only been tackled recently~\cite{Afferrante2022TI,Perez-Rafols2022JMPS}.
Perez-Rafols \textit{et al.} simulated a parabolic tip with single-wavelength roughness and found contributions of viscoelasticity and waviness to adhesion hysteresis
to be nearly independent and additive as long as the viscoelasticity was confined to the edges of the wavy contact. 
However, despite being cutting edge, the study lacks comparison to experiments and is limited a single relaxation time and one-dimensional interfaces. 
{
Violano \textit{et. al.}~\cite{Violano2021MM, Violano2021MMa} also modeled two surface dimensions with a bearing-area model and successfully compared to experiments.
However, the surfaces were designed to have random asperity heights without spatial correlation, which formally corresponds to a Hurst exponent of $H=-2$.
For $H\le -1$, scaling laws deduced from rigorous simulations~\cite{Campana2008JPCM} or Persson theory~\cite{Persson2008JPCM} predict spatial correlations to be absent at large distance so that one can get away with bearing-area models in this atypical situation.
}
%

The central difficulty when conducting {rigorous, two-dimensional} simulations lies in the short-range nature of adhesion, whose range of interaction $\rho$ critically affects not only the viscoelastic losses caused by propagating cracks~\cite{Mueser2022EL} but also the energy hysteresis induced by elastic instabilities~\cite{Ciavarella2017JMPS,Wang2021L}.
%
Unfortunately, using realistically small values for $\rho$ requires extremely fine discretization to be used so that lattice instabilities are avoided~\cite{Wang2021L}.
The latter would lead to Coulomb friction for propagating cracks rather than to the more realistic polynomial crack-speed dependence~\cite{Schapery1975IJF,Persson2005PRE}.
%
%
%
As of now, it does not seem to be clear how to reproduce reliably realistic dynamics of viscoelastic adhesion theory with continuum-theory based simulations.


In this work, we study the contact between a viscoelastic film and a nominally flat, cylindrical punch to which single-wavelength, small-scale roughness is added.
Depending on the relative orientation of different wavevectors $\vb{q}$, which all have the same magnitude $q$, different patterns can be produced for which the local height maxima form either a hexagonal or a triangular lattice. 
The questions to be addressed in this study are manifold.
Can simulations reproduce experimentally observed dependencies, such as the normal force as a function of time and the concomitant contact-area evolution?
How does the unit of time, or retraction velocity, have to be renormalized for a successful comparison between simulation and experiment when it is computationally unfeasible to work with realistically small values of $\rho$?
Is it possible to clearly discriminate between dissipation due to elastic instabilities and viscoelastic crack propagation? 
And last but not least, can visualizing the contact area aid the prediction of imminent contact failure?
{The latter question can be relevant for modern adhesive gripping devices coupled with machine learning and robotics for performance prediction and automation \cite{Tinnemann2019AFM,Samri2022MT}.}

The remainder of this paper is organized as follows:
\cref{sec:MoMe} summarizes the ideal reference model, the computational approach, and the experimental methods.
Results are presented in \cref{sec:results}.
A detailed discussion is given and conclusions are drawn in \cref{sec:discConc}.

\section{Models and methods}
\label{sec:MoMe}

\subsection{Reference model}
\label{sec:refMod}

\begin{figure}[hbtp]
\centering
\includegraphics[height=1.5in]{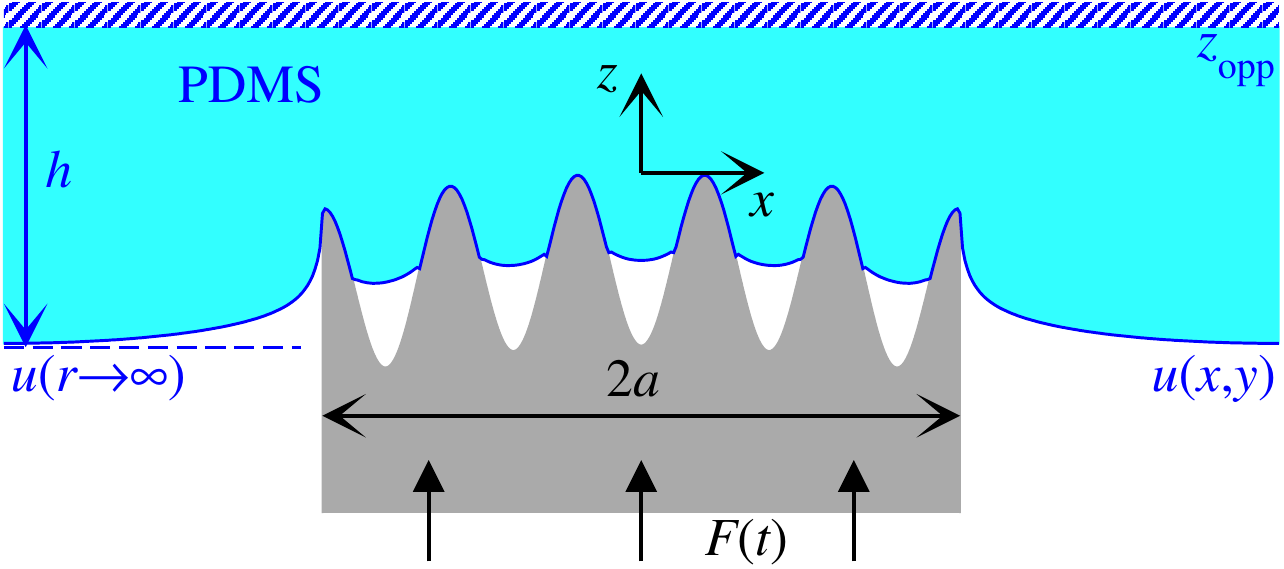}
\caption{Illustration of the reference system.
{Normal and lateral} dimensions are not to scale. However, $u(x,y)$ and the indenter shape represent  data obtained from the simulation during compression. {The used height profile reflects deviations from the target and the true sinusoidal undulations of the flat punch.} 
\label{fig:model}
}
\end{figure}

In this work, we compare simulations and experiments mimicking an ideal (mathematical) reference model, which is sketched in \cref{fig:model}. 
It consists of a flat, cylindrical, perfectly rigid punch of radius $a$ to which single-wavelength corrugation $z(x,y)$ is added.
The punch is indented into a homogeneous, isotropic, and elastomeric film with linear viscoelasticity.
Inspired by the experimental realization, we will call this material PDMS, although the theoretical model does not necessarily imply a specific polymer compound.
{The elastomer} has a finite height $h$, infinite in-plane dimension with a frequency-dependent Young's modulus $E(\omega)$ and a constant Poisson's ratio $\nu$.
``PDMS'' and punch interact through a cohesive-zone model, which is characterized by a surface energy per unit area $\gamma$ and a small but finite interaction range $\rho$.
Punch and elastomer are frictionless and cannot interpenetrate.

Numerical values of the reference model are
$a = 375~\textrm{µm}$, $h = 2~\textrm{mm}$, $E(0)=2~\textrm{MPa}$, $E(\infty) = 2~\textrm{GPa}$, $\nu = 0.495$, and $\gamma = 50~\textrm{mJ/m}^2$, which are admittedly our best guesses for the values of the laboratory version of the reference model. 
The precise frequency dependence of $E(\omega)$ as well as the interaction range cannot be well matched between the laboratory and the \textit{in-silico} realization of the reference model, which is why we abstain from defining reference values here.
The experimental range of adhesion can certainly be classified as short-ranged, while that used in the simulations is merely as short-ranged as computationally feasible. 

\begin{figure}[hbtp]
    \centering
    \includegraphics{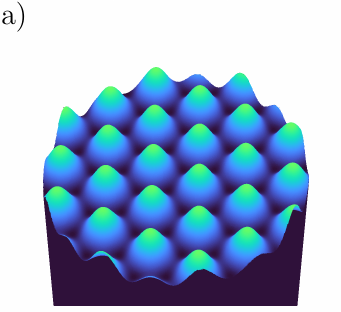}
    \includegraphics{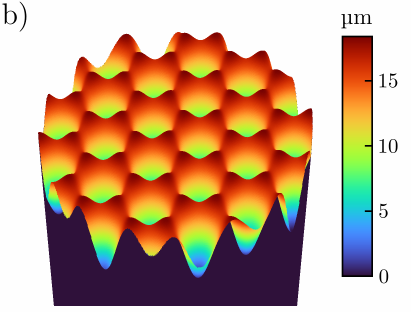}
    \caption{Top view of the flat indenter with (a) triangular and (b) hexagonal waviness.
     Height and lateral dimensions are not to scale.}
    \label{fig:surfs}
\end{figure}

Two different height topographies are added to the punch, a triangular (tri) and a hexagonal (hex) one.
Redefining prefactors compared to previous work~\cite{Dapp2015EL}, they are given by  
\begin{subequations}
\begin{align}
    \frac{z_\textrm{hex}(x,y - \sqrt{3}\lambda/4)}{z_0(\textrm{hex})} & =  \frac{4}{9}\qty{\frac{3}{2}+2\cos\qty(qx)\cos\qty(\frac{1}{\sqrt{3}}qy)+\cos\qty(\frac{2}{\sqrt{3}}qy)}\\
    \frac{z_\textrm{tri}(x,y)}{z_0(\textrm{tri})} & = 2 - \frac{z_\textrm{hex}(x,y)}{z_0(\textrm{hex})},
\end{align}
\end{subequations}
where $q = 2\pi/\lambda$ is the wave vector and $\lambda = {150~\textrm{µm}}$.
The amplitude of the undulations---defined as half the difference between maximum and minimum---are set to 
$z_0(\textrm{hex}) = {9.2~\textrm{µm}}$ and
$z_0(\textrm{tri}) = {4~\textrm{µm}}$.
Resulting punch profiles are shown in \cref{fig:surfs}.
Different amplitudes were chosen, because the jump into contact of saddle points occurs much earlier for hexagonal than for triangular corrugations~\cite{Dapp2015EL}. 
With these choices of $z_0$, the radii of curvature of the asperities turned out to be {$R_\textrm{c} \approx {150~\textrm{µm}}$} for both profiles.
Moreover, the dimensionless surface energy $\tilde{\gamma} \equiv \gamma/v_\textrm{ela}^\textrm{full}$, where $v_\textrm{ela}^\textrm{full}$ is the 
 areal elastic energy in full, static contact, $v_\textrm{ela}^\textrm{full}$,  are approximately $\tilde{\gamma}(\textrm{tri}) \approx 0.32$ and $\tilde{\gamma}(\textrm{hex}) \approx 0.061$ {for the respective, periodically repeated wave patterns}.
 These values are less than $1/2$, which has been identified as the (approximate) dividing line between sticky and non-sticky for many surfaces with a symmetric height distribution~\cite{Wang2022FME}.

The indenter is moved from non-contact at different constant velocities $v_\textrm{ext}$
ranging from 0.5 to ${25~\textrm{µm/s}}$ into the elastomer until a target force, or preload, $F_\textrm{pl}$, is reached, at which point the velocity is reverted quasi-instantaneously to initiate detachment. 
The preload is varied between 1 and {10~\textrm{mN}} for the hexagonal and between 40 and {80~\textrm{mN}} for the triangular surface.

A brief note on the choice of the frequency- and wavenumber-independent Poisson's ratio is in order.
Real elastomers deviate from ideal incompressibility at high frequency much more than at low frequency, i.e., their Poisson's ratio falls from just below 0.5 at $\omega \to 0$ to typically around 0.3 for large $\omega$~\cite{Caracciolo1996MM,Tschoegl2002MTM}. 
In the present study, we can ignore this effect, because the film thickness clearly exceeds the punch radius, which means that all relevant modes, other than the center-of-mass mode, can be treated as if the film was semi-infinite. 
In this case, the contact modulus, $E^*(\omega) = E(\omega)/\qty{1-\nu^2(\omega)}$, which is not very sensitive to the frequency dependence of the Poisson's ratio, becomes the central elastic parameter determining the viscoelastic response.

\subsection{Numerical model and methods}
\label{sec:methNum}

The solution of the dynamics defined implicitly in \cref{sec:refMod} requires some idealizations to be given up, while other specifications can be perfectly realized, at least to numerical precision. 
The latter include linear elasticity, the  topographies, velocities, loads, and any other specified number.
Compromises are related to the numerical solution of the problem, which include the necessity to discretize space and time as well as the use of periodic boundary conditions for reasons of efficiency. 

\subsubsection{Reproducing viscoelastic properties using GFMD}

\begin{figure}[hbtp]
    \centering
    \includegraphics[width=3.5in]{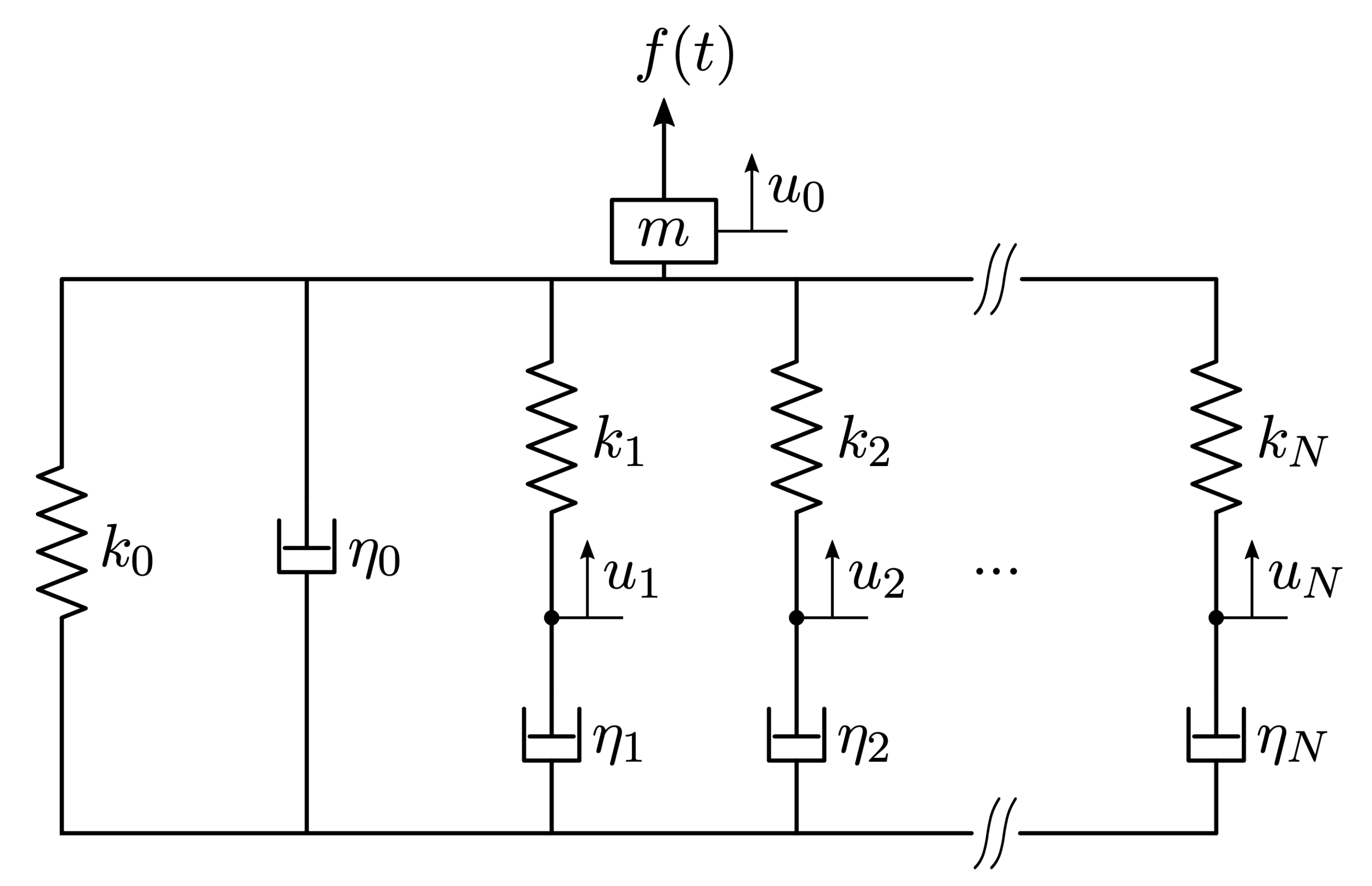}

    \caption{Illustration of the rheological model employed, which consists of one Kelvin-Voigt element $(k_0, \eta_0)$ and $N$ Maxwell elements $ (k_n,\eta_n)$ in parallel plus an inertial mass $m$.  In GFMD, each  mode $\tilde{u}(\vb{q}) \widehat{=} u_0$ is represented with such a model.  
    }
    \label{fig:GMWmodel}
\end{figure}

\begin{figure}[hbtp]
    \centering
    \includegraphics[scale=1]{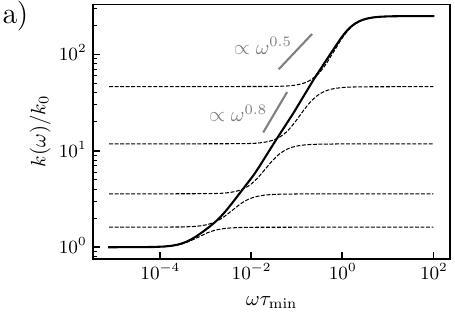}
    \includegraphics[scale=1]{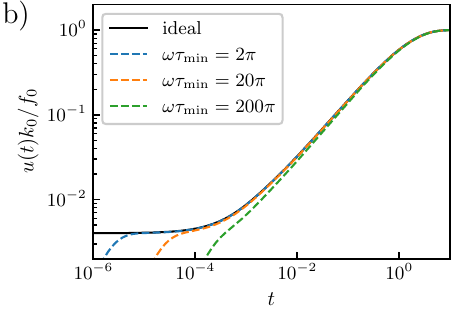}
    \caption{
    (a) Frequency-dependent target stiffness $k(\omega)$ as a function of frequency $\omega$ using {$k_{n+1} = 6^{0.8} k_n$}, $\tau_{n+1} = \tau_n/6$, and $N=5$.
    Dotted lines show the relaxation process of individual Maxwell elements and the solid, gray line a $\omega^{0.8}$ power law.
    {A more realistic $\omega^{0.5}$ power law is shown for comparison.}
    (b) Associated response function $u(t)$ to a point force $f(t) = f_0 \Theta(t)$ using different auxiliary masses leading to different eigenfrequencies $\omega_\textrm{GFMD} = \sqrt{k_\infty/m}$.
    {The full black line shows the ideal or target response function, while dashed colored lines reflect the implemented response function.}
    In each case, the auxiliary damping was chosen to satisfy the condition for critical damping $\eta_0 = 2m \omega_\textrm{GFMD} $.
    }
    \label{fig:maxwellStuff}
\end{figure}

The time evolution of {an isotropic and linearly} elastic bottom layer,   {being infinitely large in the plane or periodically repeated}, can be cast as
\begin{equation}
    \tilde{u}(\vb{q},t) = \int\limits_{-\infty}^{t} 
    \dd t' \, \tilde{G}(q,t-t')\, \tilde{f}(\vb{q},t'),
\end{equation}
{for reasons of symmetry}.
Here, $\tilde{u}(q,t)$ is the spatial Fourier transform of the displacement field as a function of time $t$, $\tilde{f}(\vb{q},t)$ is the spatial Fourier transform of the external force per unit area acting on the elastomer, and $\tilde{G}(q,t-t')$ is the Green's function conveying the effect that this {Fourier transform}, at time $t'\le t$ has on {(the Fourier transform of)} the displacement at time $t$. 
{For a half space}, $\tilde{G}(q,t)$ is formally given by
\begin{equation}
    \tilde{G}(q,t) = \frac{2}{q} \int\limits_{-\infty}^\infty \dd\omega \frac{1}{E^*(\omega)} e^{\iu \omega t}.
\end{equation} 
The time dependence of the Green's functions $\tilde{G}(q,t)$ or the response functions they produce can be represented via a Prony series, which in turn can be realized through rheological models, as that depicted in \cref{fig:GMWmodel}, where  stiffness  ($k_n$) and damping  ($\eta_n$) terms are introduced. 
An appropriate choice of weights $\kappa_n=k_n/k_0$ and relaxation times $\tau_n=\eta_n/k_n$ allow the target frequency dependence $\kappa(\omega) = E(\omega)/E(0)$ to be approximated through
\begin{equation}
\label{eq:freqMod}
    \kappa(\omega) 
    = 1   
    + \sum_{n=1}^N \kappa_n \qty{\frac{\omega^2 \tau_n^2}{1 + \omega^2 \tau_n^2}
    + \iu  
    \frac{\omega \tau_n }{1 + \omega^2 \tau_n^2}}.
\end{equation}
An example of a system producing such a target dependence is shown in \cref{fig:maxwellStuff}a.

An inertia $m$ and damping $\eta_0$ were added to the rheological elements, which allowed us to implement the final rheological model into a Green's function molecular dynamics (GFMD)~\cite{Campana2006PRB} based code.
The two added elements alter the frequency dependence to
\begin{equation}
    \kappa_\textrm{GFMD}(\omega) = \kappa(\omega) - \omega^2 \frac{m}{k_0} + \iu \omega \frac{\eta_0}{k_0}.
\end{equation}
By replacing $k_0$ with $k_0(q) = qE^*/2$ for each $\tilde{u}(\vb{q},t)$, all $k_n$ and $\eta_n$ turn into $k_n(q)$ and $\eta_n(q)$, with the exception of $\eta_0(q)$, whose parametrization will be discussed separately.

%

The resulting equations of motion for each mode and its associated extra degrees of freedom $u_n(\vb{q},t)$ read:
\begin{subequations}
\begin{align}
    m({q}) \ddot{\tilde{u}}(\vb{q},t) + \eta_0(q) \dot{\tilde{u}}(\vb{q},t) + 
    k_\infty(q) \tilde{u}(\vb{q},t)
    & =
    \tilde{f}(\vb{q},t) + \sum_{n=1}^N k_n(q) u_n(\vb{q},t), \label{eq:MaxwellEOMa}\\
    \eta_n(q) \dot{u}_n(\vb{q},t)  
    &=  
    k_n(q) \qty{ \tilde{u}(\vb{q},t) - u_n(\vb{q},t) },
    \quad n \in 1...N,
    \label{eq:MaxwellEOMb}
\end{align}
\end{subequations}
with $k_\infty(q) \equiv \sum_{n=0}^N k_n(q)$, which is nothing but $k_\infty(q) = E(\infty) k_0(q)/E(0)$, where only one of the two ``arrays'' $k_0(\mathbf{q})$ and $k_\infty(\mathbf{q})$ needs to be stored in memory.
Even for a single Maxwell element, the solution of the equations of motion turned out simpler and more stable (but not necessarily faster) than our previous extension to GFMD~\cite{Sukhomlinov2021ASSA}, which was similar in spirit to that proposed by van Dokkum and Nicola~\cite{vanDokkum2019MSMSE} in that the first-order time derivatives of the external forces were needed. 
Our current approach  rather resembles that pursued by Bugnicourt \textit{et al.}~\cite{Bugnicourt2017TI}, who used Zener instead of Maxwell models and a conjugate gradient (CG) minimization method for the solution of the  instantaneous or high-frequency response instead of the auxiliary masses. 
%

Before proceeding, a few additional notes of clarification might be in order.
First, tildes on the $u_n(\vb{q},t)$ are omitted, as they are not subjected to an inverse Fourier transform. 
Second, the equations of motion solved in conventional GFMD are recuperated by setting $N=0$, while the standard linear solid is obtained when using $N=1$ and (infinitesimally) small values for $m$ and $\eta_0$. 
Third, the presented methodology is readily extended to more general situations, even if the above treatment merely targets the specialized problem defined in \cref{sec:MoMe}.
For example, if the elastic properties were anisotropic in the $xy$-plane, as they would be if the elastomer were prestrained in $x$ but not in $y$ direction, the coefficients $k_0(q)$ and thereby $k_n(q)$ and $\eta_n(q)$ would be functions of the vector $\vb{q}$ and not merely of its amplitude.
Similarly, if the elastic properties changed with depth, as is the case when the crosslinking and thus the stiffness depends on the depth~\cite{Mueser2019L}, but similarly when the elastomer is confined by a hard wall~\cite{Carbone2008JMPS,Carbone2009EPJE}, the term $k_0(q) = qE^*/2$ would have to be replaced or multiplied with an appropriate $q$-dependent function.
Last but not least, using $N$ Maxwell elements does not imply a single time step to take $N$ times longer than a conventional GFMD time step, because the most demanding operation is the fast Fourier transform.
For example, using $N=5$ Maxwell elements per mode only increases the CPU time per time step by roughly 50\%, compared to a regular GFMD time step for a discretization of $2,048\times 2,048$. 
Relative costs on memory are clearly larger.
The reason why we do not go beyond five Maxwell elements in this study is that almost four decades of relaxation times can be covered when choosing $\tau_{n+1}=\tau_n/6$, which requires the time step $\Delta t$ to be chosen very small assuming $\tau_1$ to remain fixed. 
Mimicking an even broader relaxation-time spectrum would impose further and eventually unfeasible demands on the used time step $\Delta t$. 

While the values of $k_0(q)$ as well as $k_n(q)$ and $\eta_n(q)$ for $n\ge 1$ are predetermined by $\kappa_n$, $\tau_n$, 
and $E^*$, the remaining parameters $m(q)$ and $\eta_0(q)$ should be chosen such that they provide a compromise between accuracy and efficiency. 
The goal must be to find the high-frequency elastic response as quickly as possible, albeit without making it necessary to dramatically reduce $\Delta t$.
Under the made assumption that $E(\omega)$ does not depend on $q$, each free surface mode must have the same response function.
This implies  $m(q) \propto k_0(q)$, which is the choice made in so-called mass-weighted GFMD~\cite{Zhou2019PRB}.
The period associated with the resulting frequency $\omega_\textrm{GFMD} = \sqrt{k_\infty(q)/m(q)}$ is best chosen such that it is not much {larger than $1/\tau_\textrm{min} = 1/\tau_N = 1/\min(\tau_n)$.}
%
We found the ``aggressive''choice of $\omega_\textrm{GFMD} \tau_\textrm{min} = 2 \pi$ to be sufficient. %
%
If, however, the pulling velocity is so large that the time step $\Delta t$ is no longer limited by  $\tau_\textrm{min}$ but by a large pulling velocity, e.g., by the ratio of a characteristic height amplitude and the pulling velocity, we recommend to set $m$ such that $\omega_\textrm{GFMD}\Delta t  \approx  \pi/10$ as to achieve a numerically stable but fast relaxation of the high-frequency response to its exact solution.
After realizing that the left-hand side of \cref{eq:MaxwellEOMa} represents a damped harmonic oscillator, $\eta_0(q)$ is set to satisfy the condition for critical damping, i.e., $\eta_0(q) = 2m(q) \omega_\textrm{GFMD}$.

As a consequence of the just-made choices, the target viscoelastic response, for example, to an indenter exerting a force on a single (grid) point starting at time $t_0$, is mimicked quite accurately at times satisfying  $t > t_0 + \tau_\textrm{min}$, which can be achieved within one or two dozen time steps. 
The validity of this claim is demonstrated in \cref{fig:maxwellStuff}b for our system with $N=5$ Maxwell models. 
It can be seen that even $\omega_\textrm{GFMD}\tau_\textrm{min} = 2\pi$ leads to quite satisfactory results, although the time step, $\Delta t$  
was set by default to {$\Delta t = \tau_\textrm{min}/20$}.
The ratio $k_\infty/k_0$ was reduced from its reference value of 1,000 to 250 because this made the $\kappa(\omega)$ dependence at small $\omega$ be closer to the real PDMS~\cite{Tiwari2017SM} when using a single Maxwell element. 
When using five Maxwell elements, in addition to the $k_0$ spring, we can produce a response function $E'(\omega)$ that roughly scales proportional to $\omega^{\beta}$ with $\beta \approx 0.8$ at an intermediate frequency $\omega_\textrm{int}$ defined through $E'(\omega_\textrm{int})=\sqrt{E'(0)E'(\infty)}$, where $E'(\omega)$ is the storage modulus, i.e., the real part of the complex function $E(\omega)$. 
A single element yields $\beta\approx 2$, while experimental systems 
{are often close to $\beta \lesssim 0.5$~\cite{Tiwari2017SM}}. 
{An exponent of $\beta = 0.8$ thus appeared a good trade-off between computational efficiency and reality.}


\subsubsection{Modeling adhesion} 
\label{sec:modAd}
The adhesive and repulsive interaction between elastomer and indenter is modeled by the cohesive zone model (CZM) proposed in Ref.~\cite{Wang2021L}.
Assuming their two surfaces with nominal surface energy $\gamma$ to have a gap $g(x,y)$, the interaction potential $\Gamma(g)$ is given by 
\begin{equation}
\Gamma(g) = -\gamma\cdot
\begin{cases}
    \qty{1+ \cos(\pi g / \rho)} / 2  & \textrm{for }\; 0 \le g < \rho \\
    \qty{ 1 - (\pi g / \rho)^2/4 }      & \textrm{for }\; g < 0 \\
   \;  0 & \textrm{else}
\end{cases},
\end{equation}
where $\rho$ is the range of adhesion.
Our CZM allows two surfaces to overlap marginally but penalizes the overlap with a harmonic function.
Enforcing a strict non-overlap constraint might be possible, albeit only at a much enhanced numerical cost, since this would certainly require all internal modes $u_n(\mathbf{q})$ to be Fourier transformed.
Moreover, the quadratic dependence of the potentials implies an upper bound for the stiffness of the equation to be solved, thereby ensuring stable integration with an appropriately chosen time step. 
The maximum adhesive stress $\sigma_\textrm{th} = \max( \dd \Gamma / \dd g)$
that can locally occur using this model is $\gamma \pi / (2 \rho)$.

The range of adhesion is generally chosen such that it is as small as possible for a given discretization but not so small that lattice pinning and subsequent instabilities of the grid points at a propagating crack front would occur.
This can be achieved when the maximum curvature of the potential is set to approximately $0.2 q_\textrm{ref} E^* $, where $q_\textrm{ref}\equiv 2\pi n/L$, $n$ being the number of discretization points parallel to one spatial direction and $L$ the linear dimension of the periodically repeated simulation cell~\cite{Wang2021L}. 
Given a default choice of $L = {1.5~\textrm{mm}}$
and discretizations of the elastomer surface into grid points whose number ranged from $2,048 \times 2,048$ to $4,096 \times 4,096$, $\rho$ turned out to lie in between 0.187 and ${0.264~\textrm{µm}}$, which is not only much more than typical Lennard-Jones interaction ranges of ${3~\textrm{Å}}$ but also exceeds recent estimates~\cite{Thimons2021EM}, which were obtained from experimentally measured pull-off forces between ruby and diamond, by a little more than a factor of ten.

To meaningfully compare simulations and experiments, it is necessary to assess whether the adhesive interactions used in the model are short- or long-ranged.
This can be done using a (generalized) Tabor parameter, which is defined as the ratio $\mu_\textrm{T} = \rho_\textrm{c}/\rho$, where  $\rho_\textrm{c}$ is a characteristic interaction range at which the cross-over from short- to long-ranged adhesion takes place.
Assuming that $\gamma/E^*$ and a characteristic radius $R_\textrm{c}$ are the only two independent length scales that can be constructed from the model, the only possible dependence of $\mu_\textrm{T}$ on the two length scales is
\begin{equation}
\label{eq:Tabor}
    \mu_\textrm{T} = \frac{1}{\rho} R_\textrm{c}^\beta \left(\frac{\gamma} {E^*}\right)^{1-\beta},
\end{equation}
assuming either a flat punch with radius $R_\textrm{c}$ or an indenter whose shape is a power law in the radius, i.e., $h(r) = R_\textrm{c} (r/R_\textrm{c})^n/n$.
It will be shown in a separate work that the exponent $\beta$ turns out to be $\beta = (n-1)/(2n-1)$ so that $\beta = 1/3$ for a parabolic ($n=2$) and $ \beta = 1/2$ a flat-punch ($n\to\infty$) indenter. 
These two limiting cases agree with the definition of the conventional Tabor parameter for a parabolic indenter~\cite{Tabor1977JCIS} and for the parameter allowing one to assess if the high-velocity retraction of a flat-punch indenter fails through crack propagation or through uniform bond breaking.
They correspond to the limits of $\mu_\textrm{T} \gg 1$ and $\mu_\textrm{T}\ll 1$, where the  high-frequency rather than the small-frequency modulus is used in the calculation of the Tabor parameter~\cite{Persson2003W}. 

The numerical Tabor parameters at the scale of local parameters turns out to be $\mu_\textrm{T} \approx 2$ for either profile when using the default discretization of $4,096\times 4,096$ and thus $\rho={0.187~\textrm{µm}}$.  
{This is because the radii of curvature associated with the peaks of the (ideal) profiles have similar values, namely $R_\textrm{c} = {163~\textrm{µm}}$ (triangular)and $R_\textrm{c} = {142~\textrm{µm}}$ (hexagonal).}
%
%
While $\mu_\textrm{T} \approx {2}$ produces (quasi-static) load-displacement curves in contact similar to short-range adhesion~\cite{Mueser2014BJN,Wang2021L}, it must be considered long-ranged in non-contact~\cite{Ciavarella2017JMPS,Wang2021L}.
{This is because the jump-into contact occurs at a relatively large separation so that the adhesion hysteresis is about 50\% of the true hysteresis for parabolic indenters with $\mu_\textrm{T} = 4$.
From that point on, adhesion hysteresis converges only with the cubic root of the linear mesh size to the exact result~\cite{Wang2021L}}.
Consequently, simulations cannot be expected to reproduce experimental results with close-to-perfect precision, at least not using currently available methods and computers. 
If surfaces were not corrugated, the generalized Tabor parameter for the flat punch would be reasonably large, i.e., $\mu_\textrm{T} \approx 10$ for the $2,048 \times 2,048$ resolution and $\mu_\textrm{T} \approx 14$ for  $4,096 \times 4,096$.
It may also be of interest to calculate the Tabor parameter at a coarse scale, i.e., the one that is obtained when using the measured quasi-static pull-off force (from which an effective surface energy can be constructed) and the given range of adhesion while assuming a perfectly flat punch. 
{For the hexagonal surface, these ``effective'' Tabor parameters turn out to be 0.880 and 1.24 for $2,048 \times 2,048$  and  $4,096 \times 4,096$, respectively. The triangular variant shows a 40\% smaller quasi-static pull-off force and hence an equally reduced effective Tabor parameter.}

\subsubsection{Refinements and corrections}
\label{sec:correction}

A few adjustments were made to the numerical model in order to facilitate the comparison between simulations and experiments.
Firstly, the velocity inversion was not abrupt but happened over a few but sufficiently many time steps to yield a smooth force-distance relation.
Secondly, the 3D printing process introduces deviations from the ideal reference model, most notably an undesired macroscopic curvature, which was reflected in the numerical model. 
This curvature is a result of shrinkage induced by cross-linking during UV-curing.
In selected simulations, we also accounted for the quasi-discrete height steps of $\Delta z = {0.2~\textrm{µm}}$, which result from the layer-by-layer nature of the printing process.
Final results were only marginally affected by this since $\Delta z$ is of similar order of magnitude as our interaction ranges $\rho = 0.187~\text{to}~{0.264~\textrm{µm}}$ and the steps in the topography are not very sharp.

\begin{figure}[hbtp]
    \centering
    \includegraphics{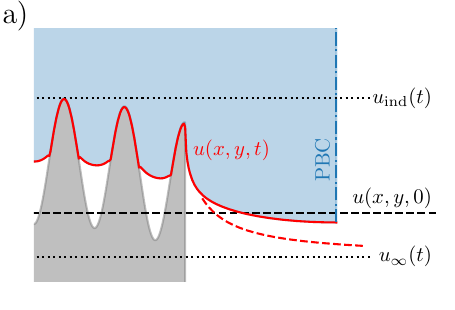}
    \includegraphics{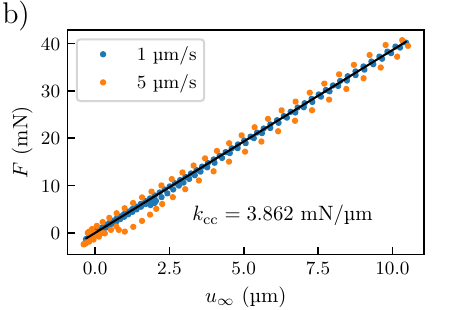}
    \caption{a) Illustration of the different displacements considered for compliance correction. b) Linear fit of $u_\infty$ for stiffness evaluation. 
    }
    \label{fig:complianceCorr}
\end{figure}

A final technical aspect deserves mentioning. 
For reasons of computational efficiency, the buffer between the indenter and its periodic image should be made as small as possible but large enough so that the stress field on the indenter is not significantly affected. 
This is achieved quite well with our choice of $L = 4a$, {supposedly because the displacement field of standard indenters approach the $1/r$ asymptotes quite closely at a distance from the symmetry axis being twice the contact radius.}
{However, the center-of-mass mode of a periodically repeated surface, $\tilde{u}(q=0)$}, deviates from the real $u_\infty = u(r \gg a)$ that would be obtained in a real system without periodic boundary conditions (PBC) {and with respect to which the indenter penetration is measured}.
An example for this difference is depicted in the form of the dashed and solid red lines in \cref{fig:complianceCorr}a.  
Given that $a/h=5$ yields a contact stiffness only {20\%} in excess of the semi-infinite case~\cite{Hensel2019JA,Mueller2022JA}, the system can be approximately treated as semi-infinite so that the correction
\begin{equation}
u_\infty \approx 6 u(L_x/2,L_y/2) - 5 u(L_x/2,0)
\end{equation}
can be used, which was originally identified for sharp indenters in square simulation cells~\cite{Mueser2014BJN}.

The described adjustment can also be thought of as a correction of an unwanted finite stiffness in the system, which does not always require a change of the experimental/numerical procedure. 
If the mismatch between ideal and measured indenter penetration, $u_\textrm{ideal}$ and $u_\textrm{ind}$, is caused by a quasi-static elastic stiffness $k_\textrm{cc}$, it can be accounted for by {adding the missing displacement} during data post-processing with the correction
%
{
$u_\textrm{ideal}(t) \approx u_\textrm{ind}(t) + \Delta u_\textrm{cc}$ with $\Delta u_\textrm{cc} = F(t)/k_\textrm{cc}$ 
and $k_\textrm{cc}$ determined from $u_\textrm{ideal}(t) - u_\textrm{ind}(t) = u_\infty(t) = F(t)/k_\textrm{cc}$ as shown in \cref{fig:complianceCorr}b.
Similarly, the experimental curves must be corrected for the machine compliance or machine stiffness $k_\textrm{M}$ to $u_\textrm{ideal} = u_\textrm{ind}(t) - F(t)/k_\textrm{M}$.
It turned out that the visualization of differences between experiments and simulations was best when subtracting the correction $\Delta u_\textrm{cc}$ from the experiments rather than adding it to the simulations. 
}

%

An issue to keep in mind regarding compliance corrections is that the occurrence of local instabilities, e.g. pull-off events, generally depends on the global system compliance.
As such, the effect of $k_\textrm{cc}$ on the load-displacement curves cannot be rigorously accounted for in post-processing~\cite{Booth2021APL,Hensel2021AAMI}.
For viscoelastic systems, it may also happen that a simulation or measurement performed at constant speed $u_\textrm{ind}(t) = v_\textrm{ext} t$ implies that $\dd u_\textrm{ideal}(t) / \dd t$ is not exactly constant but varies over time as $(1/k_\textrm{cc}) \dd F(t) / \dd t$ or $(-1/k_\textrm{M}) \dd F(t) / \dd t$.

\subsection{Experimental methods}
\label{sec:methExp}

\begin{figure}[hbtp]
    \centering
    \includegraphics[scale=0.8]{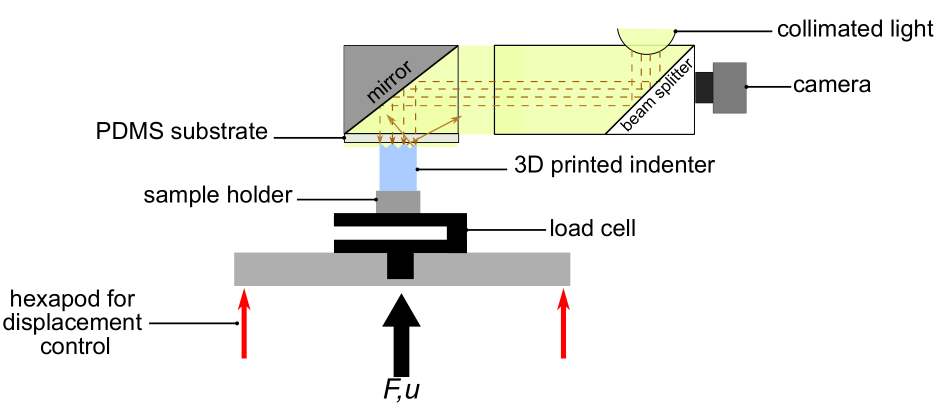}
    \caption{Schematic representation of the experimental setup.}
    \label{fig:itool}
\end{figure}

The development of optical observation techniques has benefited a wide range of applications, notably for assessing the true contact area between solids. 
Frustrated total internal reflection (FTIR) started to be applied to image the contact in the 1960s \cite{Harrick1962JAP,McCutchen1964RSI}. 
{FTIR and related methods} are routinely employed nowadays to measure stress distributions~\cite{Eason2015BB}, contact area of rough surfaces~\cite{Bennett2017TL,McGhee2017TL}, or to  visualize the contact formation and separation of fibrillar microstructures~\cite{Tinnemann2019AFM,Samri2021JAMTA,Thiemecke2020AFM,Samri2022MT,Booth2021APL}.
Despite the successful use {of FTIR} to determine multiple contact properties, obtaining high contrasts is limited to observing the contact of an opaque specimen through a transparent counter-surface.
Another technique that was employed for contact measurement is the optical interference observed as Newton’s rings \cite{McCutchen1964RSI,Wahl2008MB, Sawyer2008MB}. 
This technique became more and more relevant in contact mechanics and tribology after Krick \textit{et al.}~\cite{Krick2012TL} employed it to develop an \textit{in-situ} optical micro tribometer, which allowed them to visualize the intimate contact between solids during loading and sliding experiments. 
In this work we use a new approach for contact observation based on the coaxial lighting principle, as illustrated in \cref{fig:itool}. 
Using light from a collimated light source (collimated LED, Thorlabs, New Jersey, USA), a parallel light beam is created for homogenous lighting. 
The parallel beam is scattered at the contact points between indenter and substrate, reducing the intensity that is reflected back to the camera.
This enhances the contrast between contact and non-contact areas compared to non-parallel or transmitted light. 
It also makes it easy to keep the optics in focus, because the reflecting surface remains static during the experiment.

In preparation of the 3D printing process, the computer-generated topographies shown in \cref{fig:surfs} were converted to STL file format, vertically sliced into slabs with an adaptive thickness of 0.2 to {1.0~\textrm{µm}} and laterally hatched with a fixed width of {0.5~\textrm{µm}}. 
The resulting models were then printed by a two-photon lithography direct laser writing device (Photonic Professional GT2, Nanoscribe, Karlsruhe, Germany), using a 25x objective, writing speed of {100~\textrm{mm/s}}, and a laser power of {40~\textrm{mW}}. 
The printing material was a commercial photoresist (IP-S, Nanoscribe, Karlsruhe, Germany) used in dip-in mode, with an elastic modulus of $E_\text{IP-S}={1.34~\textrm{GPa}}$. 
After being printed, the indenter topographies were measured using a confocal microscope (MarSurf CM expert, Mahr, Göttingen, Germany). 

The substrate was fabricated from PDMS (Sylgard 184, Dow, Midland, MI, USA) by mixing the base and the curing agent in a ratio of 10:1. 
The pre-polymer was degassed using a Speed-Mixer (DAC600.2 VAC-P, Hauschild Engineering, Hamm, Germany) with {2350~\textrm{rpm}} at {1~\textrm{mbar}} for {3~\textrm{min}} and then cured at {95~\textrm{°C}} for {1~\textrm{h}}. 
\cref{fig:itool} shows the employed custom setup for tack tests with optical contact imaging. 
The normal displacement was controlled by a SMARPOD hexapod (SmarAct, Oldenbug Germany) and the force was measured by a  {2~\textrm{N}} load cell.
The PDMS substrate was glued to the bottom of a transparent sample holder containing a mirror allowing the side-mounted optical system to see through.
This holder was mounted to a modular positioning system with six degrees of freedom (SmarAct, Oldenbug Germany) for precise surface alignment using two side-view cameras.
The whole mechanical setup was measured to have an effective machine stiffness of $k_\text{M} = {38.1~\textrm{kN/m}}$.
Videos of the contact evolution during the tack tests were recorded at 50 frames per second using a digital camera (DFK 33UX273, Imaging Source Europe GmbH, Bremen, Germany). 
All experiments were performed in a laboratory with regulated temperature of ${21\pm0.2~\textrm{°C}}$ and relative humidity at ${50\pm5~\%}$.

\section{Results}
\label{sec:results}

\subsection{3D printing}

\begin{figure}[hbtp]
    \centering
    \includegraphics{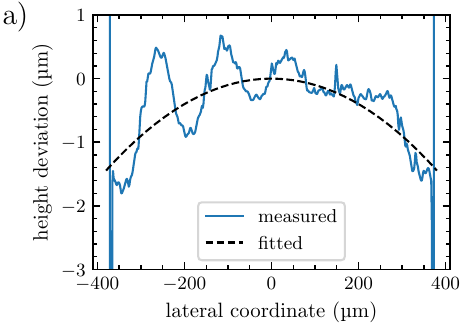}
    \includegraphics{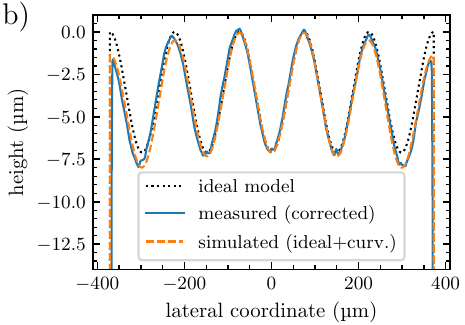}
    \caption{{a)  Deviation of an experimental line profile from the ideal model and b) absolute heights of model, simulated and experimental indenter, shown exemplarily for the triangular pattern. The curvature correction shown in a) proved necessary for a successful comparisons between simulations and experiments.}}
    \label{fig:printCurv}
\end{figure}

We first analyze optical images of the experimental topographies obtained by the 3D printing process. 
\cref{fig:printCurv}a shows the difference between  targeted and  measured height profile exemplarily for the triangular surface.
%
%
The main deviation between them is a mean curvature, which is supposedly due to shrinkage of the resin after 3D printing. 
This global curvature was reflected in the topographies used for the simulations. 
Ignoring it substantially reduces the agreement between simulations and experiment, because stresses in the flat-punch solution are largest where the correction is most noticeable.
Ideal, simulated, and experimental height profiles are compared in \cref{fig:printCurv}b.

%

\subsection{Tack tests for the triangular surface}

\begin{figure}[hbtp]
    \centering
    \includegraphics{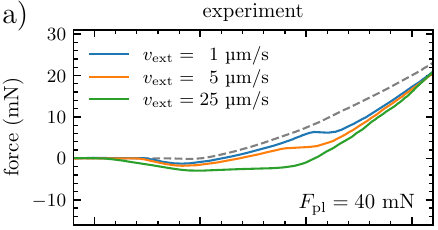}
    \includegraphics{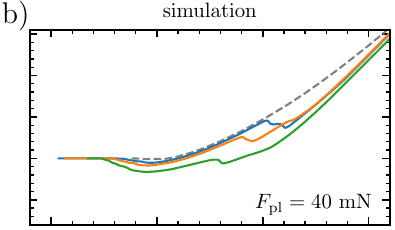}
    \includegraphics{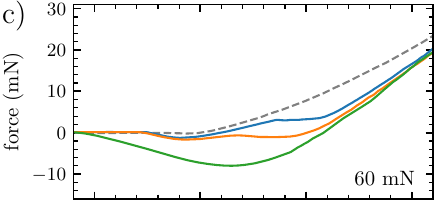}
    \includegraphics{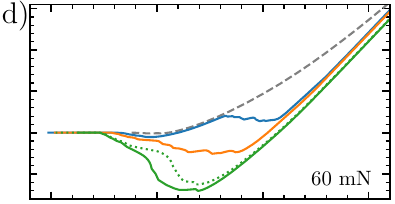}
    \includegraphics{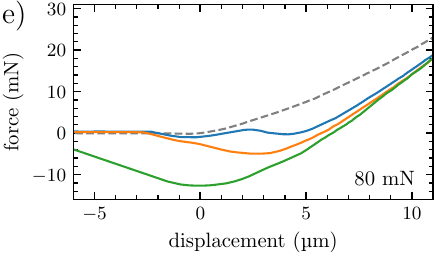}
    \includegraphics{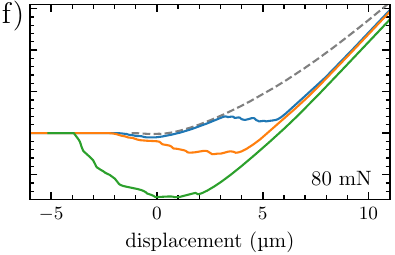}
    \caption{Load-displacement curves recorded during the detachment of the triangular surface at different velocities. The left column always shows experimental results, while the right column shows single-relaxation time simulations. From the first to the last row, the preload is increased from 40 to 60 and then ${80~\textrm{mN}}$. 
    Semi-quantitative agreement is achieved across the board, despite a slight mismatch in macroscopic contact stiffness {and the intermediate preload case}.
    }
    \label{fig:FDtri}
\end{figure}

\cref{fig:FDtri} shows the measured and simulated load-displacement curves obtained for the triangular surface.
The loading process, shown as a gray dashed line, is  smooth and rather insensitive to the approach velocity. 
The more interesting detachment parts of the curves are highlighted in color.
Experiments and simulations show similar trends:
Two bulges occur at small velocity $v_\textrm{ext}$ and small preload $F_\textrm{pl}$. 
A bulge located at slightly compressive force is related to the detachment of saddle points---as revealed in more detail further below---while the bulge at a tensile force relates to the final pull-off process.
Their locations approach each other when either $v_\textrm{ext}$ and/or $F_\textrm{pl}$ is increased.
Ultimately, they merge into a single minimum, whose value corresponds to the (negative) pull-off force. 
Although experimental and simulated curves agree only semi-quantitatively, the tensile pull-off force is increased from about $F_\textrm{po}= {2.5 \pm 0.5~\textrm{mN}}$ for a preload force of $F_\textrm{pl}= {40~\textrm{mN}}$ to {up to}
$F_\textrm{po}= {14 \pm 2~\textrm{mN}}$ for $F_\textrm{pl}= {80~\textrm{mN}}$ in both cases. 
%

{One qualitative difference between between the experimental and the simulation data is that the simulation data is more rugged. The smoother experimental data arises for three reasons: first, symmetry-related peaks in the real punch do not have the exact same height so that they depin at slightly different moments, while symmetry is perfect in simulations yielding sharper signals. Second, the real viscoelastic response function is based on a broader range of relaxation times. Third, the experimental data was low-pass filtered with a resolution of approximately {0.1~\textrm{µm}}.} 
%

{
It is quite noticeable that the comparison for $F_\textrm{pl}={60~\textrm{mN}}$ in \cref{fig:FDtri} is substantially less good than for the other two preloads.
This is because a jump into full contact occurs in the simulations a little before {60~\textrm{mN}} were reached, while valleys are a nudge short of making contact in the experiments.
One effect potentially contributing to the later, experimental jump-into-full-contact instability --- in addition to the single-relaxation time approximation --- could be the drag forces exerted by the air that needs to be squeezed out of the thin gap between elastomer and indenter before reaching full contact.}
{To indirectly account for, or, fake, the absence of full contact at intermediate loads, we added a dotted line from a simulation with a slightly reduced preload of $F_\textrm{pl}={50~\textrm{mN}}$, whereby agreement with experiments was much enhanced.
}

\begin{figure}[hbtp]
    \centering
    \includegraphics[width=5.8in]{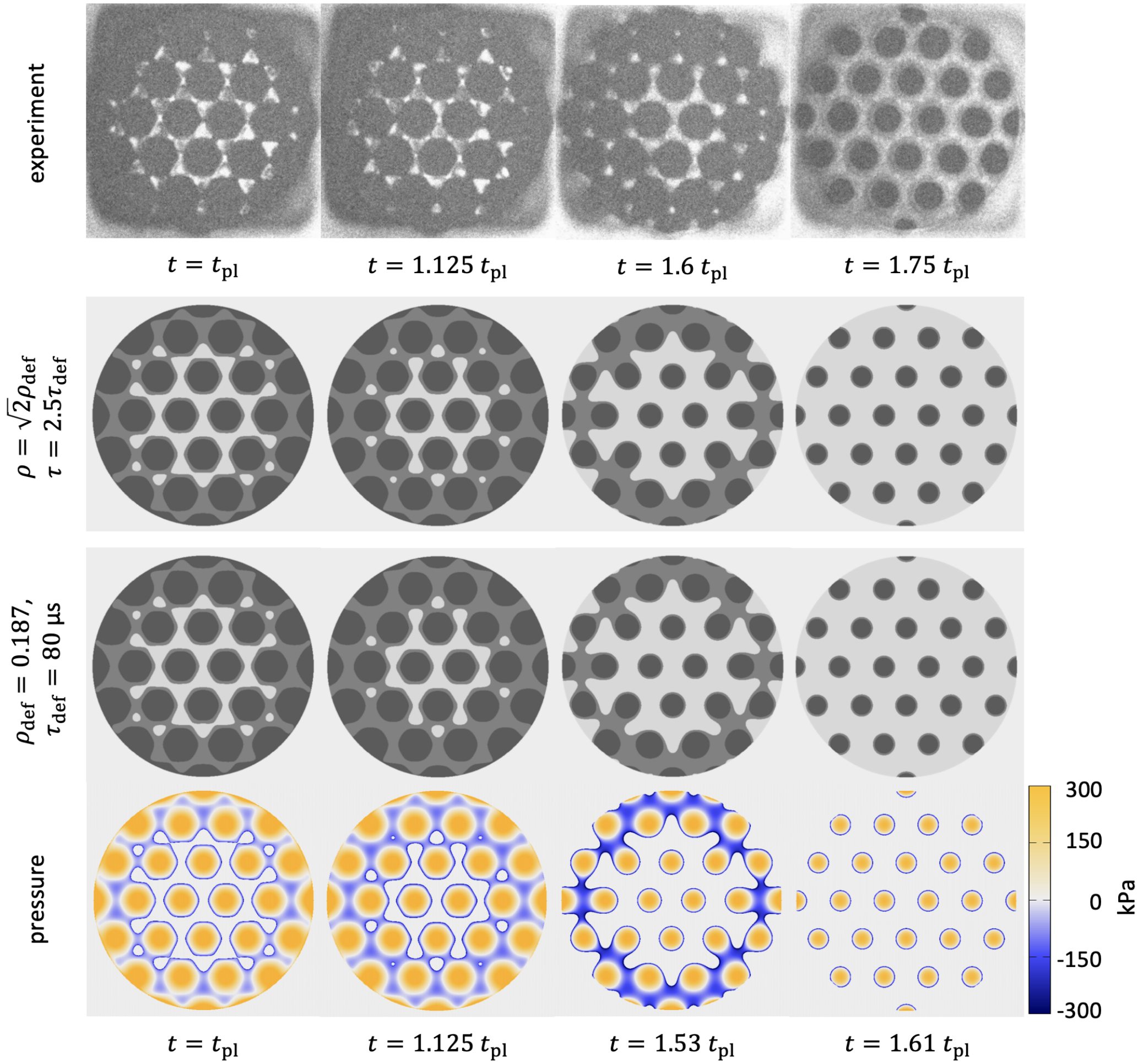}
    \caption{Contact observation during retraction in a tack test at {1~\textrm{µm/s}} and a maximum compressive load of $F_\textrm{pl} = {40~\textrm{mN}}$. The upper row shows experimental data, while the two center rows represent simulations using different relaxation time $\tau$ and range of adhesion $\rho$. The darker areas represent points with interfacial separations less than 500~nm, while the lighter areas represent larger gaps.  {The last row shows the pressure distribution associated with the previous row.} 
    $t_\textrm{pl}$ denotes the time between first contact and preload for the respective row. The times for the last two columns are located just before and past the bulge in the force-displacement curve on the compressive part of the unloading curve, i.e. at $u \approx {5~\textrm{µm}}$. Similar features are observed in all cases, e.g., the loss of contacts start with the saddle points at the edges and contact exists in all asperities but in no saddle point before the maximum tensile force during detachment is reached, i.e., at a displacement near ${{-1~\textrm{µm}}}$.
    }
    \label{fig:tri1ums}
\end{figure}

The sensitivity of the load-displacement curves w.r.t. range of adhesion and the viscoelastic model will be scrutinized after establishing that the semi-quantitative agreement between experimental and simulated load-displacement curves is not fortuitous: experimental and simulated contact topographies evolve in concert, as is revealed exemplarily in \cref{fig:tri1ums} for the preload of $F_\textrm{pl} = {40~\textrm{mN}}$ and the retraction velocity of $v_\textrm{ext} = {1~\textrm{µm/s}}$.
%
%
In the simulations representing two differently parametrized single-relaxation time models, dark gray means contact (negative gap), medium gray {is experimentally indistinguishable from contact ($0 \le g \lesssim {500~\textrm{nm}}$)}, while light gray is non-contact and very light gray the background color.
The gray shades in the real-laboratory, optical images do not allow us to determine the true interfacial separation to a high precision.
Yet, very dark pixels can be assumed to indicate contact, while less dark and bright pixels certainly imply non-contact. 
Hence, the medium-gray color level was introduced to represent gaps smaller than the medium wavelength of visible light, which we expect to appear quite dark in the optical images. 
To better visualize details of the gap distribution in the \textit{in-silico} surface, heat maps of the interfacial stresses are included in \cref{fig:tri1ums} and in later related figures for the hexagonal surface. 
%
%

Experiments and simulations reveal similar characteristics:
at the point of maximum preload, contact occurs in all peaks but only in those saddle points that are close to the outer rim, despite the slightly convex macroscopic surface curvature.
%
%
%
{The snapshots in the last two columns of \cref{fig:tri1ums} were taken right before and after the bulge in the force-displacement curve near a displacement of ${6~\textrm{µm}}$.
Hence, we can associate this bulge with the saddle-point detachment at the outer rim of the corrugated punch whenever it did occur. }

Similar qualitative agreement of the contact evolution in real-laboratory and \textit{in-silico} was found for all load-displacement curves shown in this study.
Nonetheless, quantitative differences exist:
for example, while the initial experimental and simulated frames at the maximum preload in the left column of \cref{fig:tri1ums} look astoundingly similar, given that the simulations cannot be seen as short-range adhesion on approach, the experimental contact barely changes to the next shown image. 
In contrast, the \textit{in-silico} contact reveals a noticeable retardation or aftereffect from the moments of high compression during the initial decompression in that the contact keeps growing slightly.
We attribute this to the necessity of large viscoelastic relaxation times for a proper reproduction of the dissipation caused by moving cracks. 
This makes the response to simple indentation be too sluggish so that aftereffects of the compression branch are noticeable shortly after inverting the direction of motion.
Upon further decompression, the trend reverses and the contact evolves slightly more slowly in the experiments than in the simulations: 
the destruction of contact at the saddle points
{between the last two columns of \cref{fig:tri1ums}}
happens earlier in the simulations than in the experiments.

To elucidate the role of the range of adhesion on the dynamics, we contrast the contact formation obtained in two simulations based on slightly different models, which both assume a single relaxation time and the same $E_\infty/E_0$ ratio.
The second model uses a range of adhesion that is {increased
by a factor of $\sqrt{2}$} w.r.t. the first model while the relaxation time was {multiplied with} 
2.5 to achieve close agreement between the dynamics of the two models. 
A slightly different redefinition of the relaxation time might have lead to even better agreement.
However, even with the made choice, the second and the third row of \cref{fig:tri1ums}, representing the 
{alternative}
and 
{the default }single-relaxation time model, respectively, barely allow the naked eye to distinguish the contact break-up between the two models. 
Only the second contact images, taken at a time $1.125~t_\textrm{pl}$, where $t_\textrm{pl}$ is the time elapsed between initial contact and maximum compressive load, differ slightly: {in the given time of $0.125~t_\mathrm{pl}$, the contact with the smaller relaxation time has grown more than the other one}.

The reason why changing the viscoelastic relaxation time can be ``compensated'' by a change in the range of adhesion $\rho$ during the retraction process is an interplay between the range of adhesion and the viscoelastic properties of the elastomer~\cite{Schapery1975IJF,Mueser2022EL}.
The dissipation caused by the propagating opening cracks must be reproduced in simulations in order to yield accurate load-displacement curves. 
Since steeper slopes at the contact edge imply larger (relative) velocities in a moving crack and thus enhanced dissipation, a shorter range of adhesion, leading to steeper slopes, can be compensated by shorter relaxation times used in the viscoelastic model.

\begin{figure}[hbtp]
    \centering
    \includegraphics{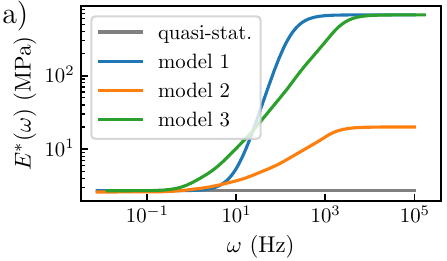}
    \includegraphics{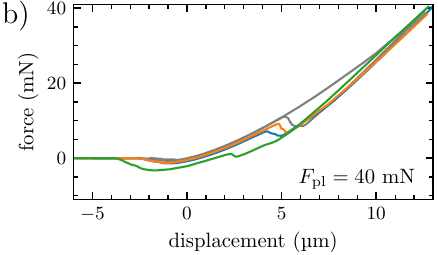} \\
    \includegraphics{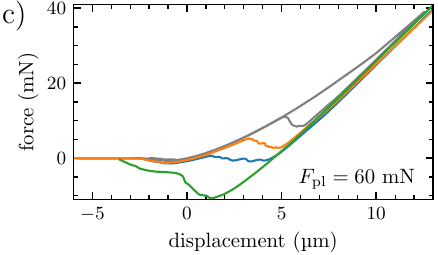} 
    \includegraphics{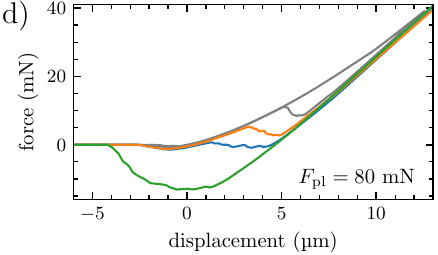} 
    \caption{Load-displacement curves obtained for the triangular surface exposed to a preload of (b) {40~\textrm{mN}}, (c) {60~\textrm{mN}} and (d) {80~\textrm{mN}} using different rheological models defined in panel (a). The detachment speed is {1~\textrm{µm/s}} in all cases.
    Larger stiffness at small frequencies leads to larger pull-off forces.
    }
    \label{fig:FDrheo}
\end{figure}

To elucidate the role of viscoelasticity, three different vicoelastic models were considered in addition to the purely elastic model reflecting the quasi-static limit. 
Their frequency-dependent contact moduli are depicted in \cref{fig:FDrheo}a with model~1 having a single relaxation time of $\tau = {{400~\textrm{µs}}}$ and $E_\infty/E_0 = 250$, while model~2 and 3 contain five relaxation times---with ratios and weight chosen as described in \cref{sec:methNum} and $\tau_\textrm{min} = {{40~\textrm{µs}}}$. 
Moreover, $E_\infty/E_0 = 8$ in model~2 and $E_\infty/E_0 = 250$ in model 3. 
Panels~(b---d) in \cref{fig:FDrheo} reveal that all three viscoelastic models increase the adhesion hysteresis with respect to the quasi-static model, which shows a rather small pull-off force of {0.7~\textrm{mN}} independent of the preload.
While the effect is relatively minor for model~2 with its relatively small $E_\infty/E_0$ ratio, the preload sensitivity is largest for model~3 with a large $E_\infty/E_0$ ratio and a tail of the ``excess''-$E(\omega)$ extending to small frequencies.
{Interestingly, the changes to the viscoelastic model in that range of frequencies seems to have a larger impact than the change associated with the high-frequency end of the spectrum.}
For the intermediate preload of {60~\textrm{mN}}, the maximum tensile force  occurs at slightly positive displacement and is clearly associated with the detachment of saddle-points rather than with that of asperity peaks.

\subsection{Tack tests for the hexagonal surface}

\begin{figure}[hbtp]
    \centering
    \includegraphics{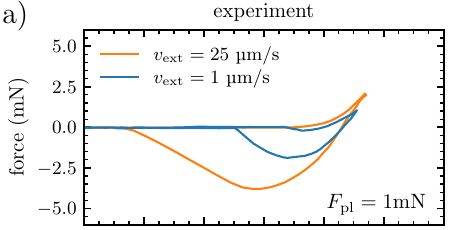}
    \includegraphics{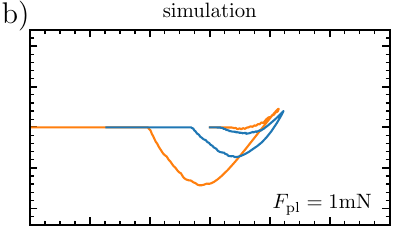}
    \includegraphics{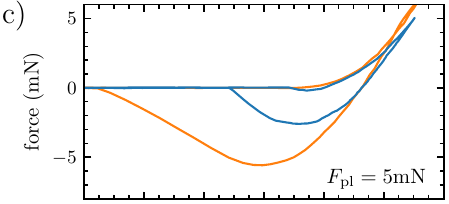}
    \includegraphics{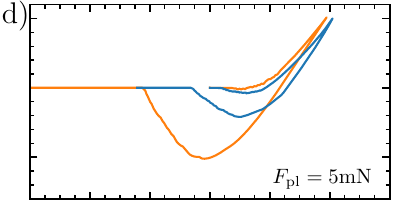}
    \includegraphics{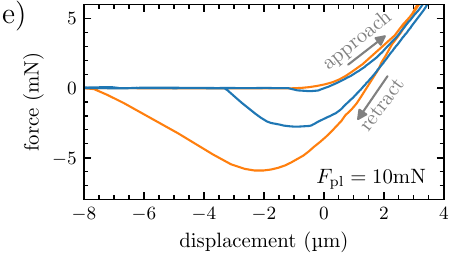}
    \includegraphics{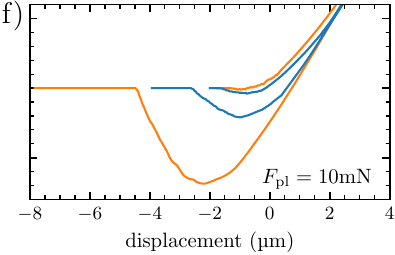}
    \caption{Load-displacement curves obtained during detachment at different velocities for the hexagonal pattern. The left column shows experimental results, while the right column shows single-relaxation-time simulations.  The preload increases from 1 to 5 and then to ${10~\textrm{mN}}$ from the first to the last row,. 
    {A quasi-static reference calculation with the same adhesive interaction resulted in a pull-off force of {1.2~\textrm{mN}} and a tensile force of {0.8~\textrm{mN}}  immediately after the jump-into-contact instability. Preload effects are distinctly reduced compared to the triangular surface in experiment and simulation alike.}
    }
    \label{fig:FDhex}
\end{figure}
\begin{figure}[hbtp]
    \centering
    \includegraphics[width=5.8in]{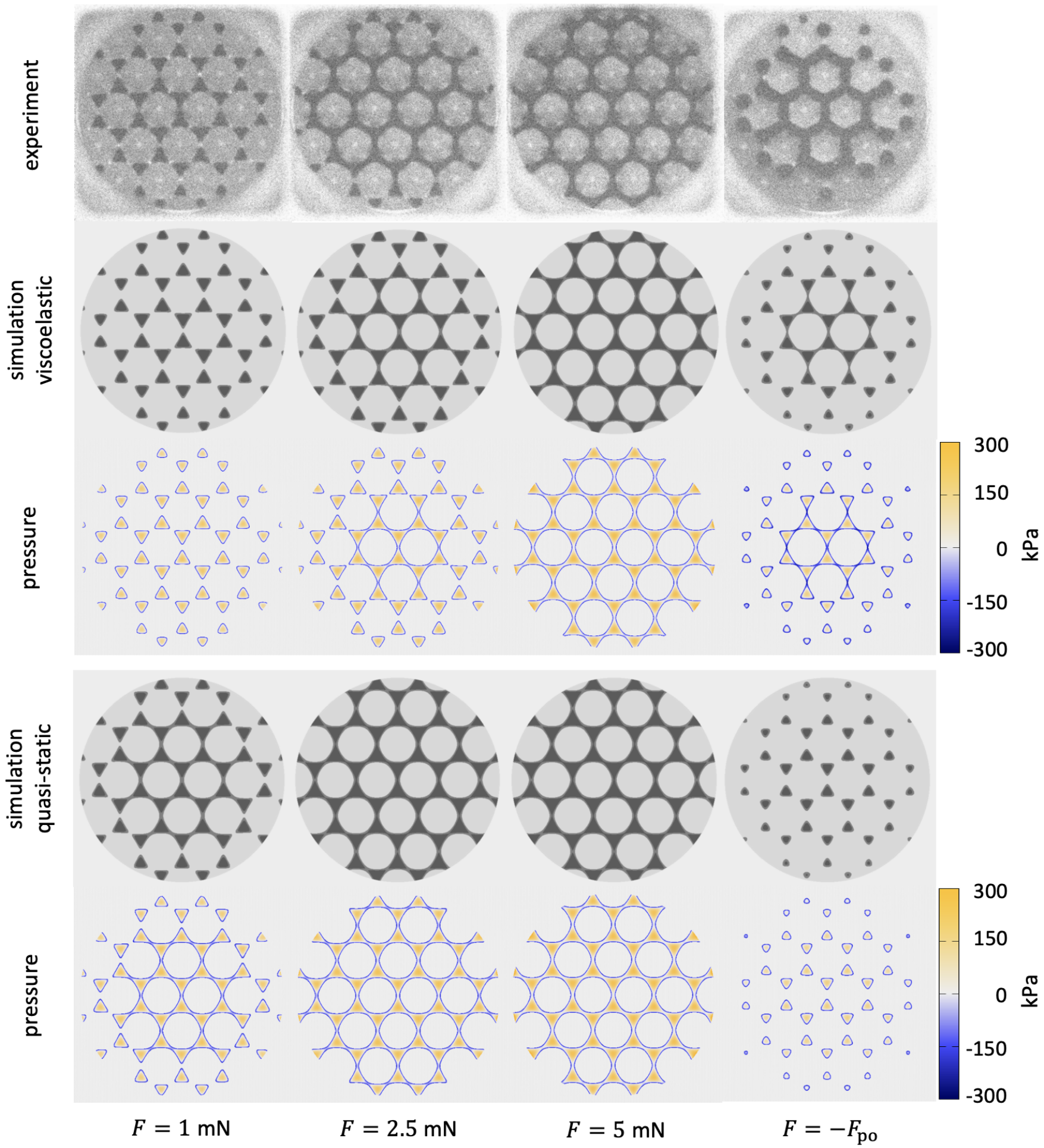}
    \caption{Contact observation during a tack test at {1~\textrm{µm/s}} and a maximum compressive load of {10~\textrm{mN}}. The upper row is taken from the experiment, the center row from a simulation with a single relaxation time ($\tau = {{200~\textrm{µs}}}$ and $\rho = {2.642~\textrm{µm}}$) and the {last two rows} from a quasi-static simulation with the same range of adhesion $\rho$. 
    Gray scales as in \cref{fig:tri1ums}.
    Frames are taken from the approach part of the tack test, except for the last ones, which reflect the moment of maximum tensile force. 
    In real-laboratory and \textit{in-silico} contacts, all maxima are always in contact while saddle points close to the rim only come into contact with increasing load. 
    The attachment of saddle points and asperities is clearly separated on approach, but their detachment occurs quasi-simultaneously.
    %
    }
    \label{fig:hex1ums}
\end{figure}

The tack tests on the hexagonal surface were carried out similarly as on the triangular surface, however using smaller preloads.
The resulting load-displacement curves are shown in \cref{fig:FDhex}, this time only for two velocities but including the loading part.
The $v_\textrm{ext} = {1~\textrm{µm/s}}$ contact evolution is depicted in \cref{fig:hex1ums} with an emphasis on the loading rather than the detachment process. 

The force-displacement curves on separation contain only one minimum at all investigated velocities for the hexagonal surface. 
The {extra} bulge related to the saddle-point detachment in the triangular surface has disappeared for the hexagonal pattern, because their detachment coincided in all investigated cases with that of the asperity peaks.
This is because saddle points are almost as high as the peaks in the hexagonal lattice. 
In fact, they are so high that contact formation of saddle points between  asperities occurs shortly after ({0.7~\textrm{µm}}) contact formation at the peaks even in the quasi-static limit on approach.
This, in turn, is due to the fact that the height of the contact line of a zero-load isolated asperity (in the Hertzian, i.e., parabolic approximation) almost extends down to a height where the corrugated profile crosses over from convex to concave. 
Due to the large dissipation of a propagating closing crack, viscoelastic saddle-point contact formation is far from being instantaneous. 
%
%
%

While the detachment curve shows fewer features for the hexagonal  than for the  triangular pattern, {the approach curves for the operating velocities of $1~\textrm{and}~{25~\textrm{µm/s}}$ no longer superimpose within line width in \cref{fig:FDhex}.
More interestingly, the load-displacement curves under compression start to be quasi-linear at a load of roughly {4~\textrm{mN}}, at which point all saddle points have made contact.
Upon decompression, the quasi-linear dependence applies to normal loads well below {4~\textrm{mN}} and ends when saddle points start to jump out of contact. 
}

Increasing the preload past the point of saddle point formation changes the load-displacement relation for the hexagonal pattern only moderately, particularly little between panels c and e of \cref{fig:FDhex}, corresponding to $F_\textrm{pl} = 5~\text{and}~{10~\textrm{mN}}$, respectively.
This can be rationalized by the contact image obtained at the maximum tensile force in the last column of \cref{fig:hex1ums}, where most saddle points are still in contact.
Those panels also corroborate the statement made at the beginning of the results section that correcting for the ``macroscopic'' surface curvature induced during cooling after the printing process was needed to achieve reasonable or, depending on viewpoint, good agreement between the laboratory and \textit{in-silico} samples:
the contact area close to the rim of the punch is noticeably reduced by the ``macroscopic'' curvature correction.

\section{Discussion and conclusions}
\label{sec:discConc}

This work addressed the interplay between viscoelastic hysteresis in contact mechanics and the hysteresis due to elastic multistability being responsible for the quasi-discontinuous snap into and out of individual contact patches observable during quasi-static driving.
To elucidate the coaction of viscoelastic and multistability effects, we studied numerically and experimentally a flat punch to which small-scale corrugation---in the form of either a hexagonal or a triangular height profile---was added.
The two height spectra are identical although the profiles are their mutual negatives, i.e., the phases of the height Fourier coefficients are shifted by $\pi$.
This makes the saddle points, which are located between two maxima and which turn out crucial for the contact mechanics, be closer to the asperity summits in the hexagonal than in the triangular lattice.

Contact of an \textit{ideal} flat punch forms quasi-instantaneously so that both viscoelastic losses due to closing cracks and multi-stability effects are negligible on approach.
Consequently, preload effects of ideal-punch detachment are minor.
However, the detachment requires a crack to propagate from the rim to the center, which leads to a viscoelasticity-enhanced work of separation at intermediate pull-off velocities~\cite{Jiang2014JPDAP,Mueser2022EL}:
the work of separation approches $2\gamma A$ at very small and very large velocities, assuming high- and low-frequency contact moduli to be well defined.

After small-scale roughness was added to the flat punch, the wavelength of the pattern being one fifth of the punch diameter, strong preload effects occurred at intermediate operating velocities but not under quasi-static driving.
Thus, preload and multi-stability effects are intertwined in the corrugated punches. 
The preload effects were distinctly larger for the triangular than for the hexagonal pattern.
%
Specifically, the pull-off force for the hexagonal lattice saturated at roughly 6 (experiment) and {7~\textrm{mN}} (simulation) once the preload had reached 5 to {10~\textrm{mN}} at an operating velocity of {25~\textrm{µm/s}}.
These two forces were roughly twice and ten times larger, respectively, for the triangular pattern.
Despite these quantitative differences, pull-off forces saturated in both cases once the preload had been large enough to induce contact at the saddle points and retraction was fast enough so that saddle-points were still in contact at the point of maximum tensile force. 
Since the saddle-point heights are rather close to (far from) the height maxima in the hexagonal (triangular) lattice, preload effects saturated earlier in the hexagonal than in the triangular system, although the hexagonal amplitude was chosen more than twice that of the triangular corrugation. 

A purely spectral approach to our system assuming random phases, as pursued in Persson's contact mechanics theory~\cite{Persson2001JCP,Persson2002PRL}, would not be in a position to reproduce or predict the observed trends.
{In the quasi-static case, the hexagonal surface pattern even shows a substantially larger pull-off force than the triangular one, despite its 2.3~times larger height amplitude.}
In principle, phase-correlation effects can be included into the theory~\cite{Muser2008PRL,Zhou2020FME}, which might fix this shortcoming. 
{Furthermore, Persson's rough surface contact theory only takes either viscoelasticity~\cite{Persson2001JCP} \textit{or} adhesion~\cite{Persson2002PRL} into account, but not (yet) both simulataneously. 
Both effects have to be accounted for in a proper description of our system.}

Can our results be rationalized with bearing-area models (BAMs), such as the popular approach by Fuller and Tabor~\cite{Fuller1975PRSLAMPS} for nominally flat, adhesive contacts? 
BAMs assume the highest asperity to come into contact first and out of contact last, the second-highest peak to come into contact second and out of contact second last, and so on and so forth.
The load-displacement laws of the individual peaks, whose shapes are approximated as paraboloids, are then added up to yield a global load-displacement curve.
While BAMs are commonly used to describe quasi-static contact loading, generalization to dynamics seems to be straightforward, e.g., by ``feeding'' the time-dependent force-displacement relation of an isolated asperity contact at the given operating velocity into the model, {see also Ref.~\cite{Violano2021MM}}. 
For our system, the radii of curvature of the hexagonal and the triangular lattice turned out quite similar. 
(The minor curvature corrections w.r.t. the ideal model changes things quantitatively but not qualitatively.) 
Thus, the depinning force of a corrugated (ideal) punch would be expected to scale linearly with the number of maxima given fixed heights and fixed radii of curvature at a fixed operating velocity.
Since the number density of maxima in the hexagonal lattice is twice that of the triangular lattice, BAMs predict roughly twice the adhesion force for our hexagonal than for our triangular patterned punch, again assuming identical velocities in both cases.
Finite-size effects and cut-off asperities at the rim of the punch renormalize that ratio but do not affect the trend.
Unfortunately, things turn out the other way around in the viscoelastic case, i.e., the triangular surface with the fewer peaks has clearly greater (viscoelastic) pull-off forces, due to the pivotal role of saddle points. 
Obviously, BAMs approximating each peak as parabolic intrinsically fail to account for saddle points, which is why we are beyond sceptical on studies reporting models in the spirit of Fuller and Tabor to be quantitative for nominally flat contacts, even if agreement can be fudged during the post-diction of experimental data.

This leaves numerical approaches, such as the here-reported number-crunching exercise, as the least problematic non-experimental tool to tackle adhesive problems similar to that investigated here.
Nonetheless, number-crunching is not entirely unproblematic either.
We also gauged the model parameters on the experiments that were reproduced, even if the few adjustable parameters were kept constant throughout all simulations.
%
One problem in the attempt to make quantitative predictions is the multi-scale nature of the dissipation during viscoelastic crack propagation.
The range of adhesion critically affects the dissipation of moving cracks, which must be reproduced correctly to model the formation and the failure of adhesive contacts reliably~\cite{Schapery1975IJF,Hui1998L,Persson2005PRE,Mueser2022EL}. 
This means that the vicinity of the crack must be resolved with a computationally unfeasible large resolution or the viscoelastic properties of the elastomer relaxation times must be rescaled, which, however, implies that the time-dependent response of the elastomer to a point indenter would no longer be correct. 
%
%
For experimental \textit{in-situ} contact observation, one challenge was to keep the focus on a moving indenter and to obtain good contrast between contact and non-contact with a lateral resolution close to the wavelength of light.  
Another difficulty was to remove artifacts from the confocal-microscopy measurements of the height profiles, which would have been even more challenging if the surfaces had had relevant roughness on finer length scales~\cite{Jacobs2022A}.

Despite all difficulties related to the numerical modeling, we would argue that the simulations matched the experiments not only qualitatively but almost quantitatively, that is, both force-distance relationships and contact images correlated quite well between simulations and experiments. 
This was accomplished not for fortuitous reasons but because (a) the simulations captured all the essential ingredients of real contacts, (b) imperfections in the 3D printing process were accounted for so that the adhesion at pull-off originated on the contact rim for the triangular pattern but in the center of the contact for the hexagonal pattern. 
{It is noteworthy that this was achievable entirely within linear response theory, neglecting in-plane stresses as well as large displacement effects.
For the observed system, we would expect these phenomena only change results quantitatively but not qualitatively as in other systems~\cite{Hui2016JPSPB,Liu2021EML}.}

{Before concluding, we would like to answer the questions raised at the end of the introduction in the order of their occurrence. First, yes, simulations can reproduce the experimental dependencies, however, at this stage, only semi-quantitatively. Second, for the current system, reducing the range of adhesion by a factor of $\sqrt{2}$ required the relaxation times to be divided by approximately $2.5$ to yield similar dynamics. This scaling might differ for other geometries. Third, we find that the dissipation due to elastic and viscoelastic instabilities cannot generally be discriminated, due to their coaction. And last but not least, contact visualization can certainly help to determine if a contact is about to break, e.g., loss of saddle points in the given system indicates imminent detachment at small retraction velocity. However, details may differ from system to system.}

Due to the good correlation between experimental and simulation results, we are confident that any (qualitative) conclusion drawn in this work is on solid grounds. 
This makes us hopeful that simulations like the ones presented here will soon be in a position to address systems beyond the demonstrator model considered here, such as pressure-sensitive adhesives or hydraulic seals in contact with surfaces having complex and not only single-sinusoidal micro-scale roughness. 
Likewise, optically studying the time evolution of contacts, in particular their saddle points, as done in this work, bears much promise to predict if a given contact is close to detachment.

\section{Declaration of competing interest}
The authors declare no conflict of interest.

\section*{Acknowledgements}
MHM thanks Anle Wang for useful discussions.
CM would like to thank Bo Persson and Leonid Dorogin for providing him with experimental DMA data for PDMS and Xuan Zhang for fabricating the model surfaces.
We acknowledge funding by the Leibniz Competition Grant MUSIGAND (No. K279/2019).


\end{document}